\documentclass[twocolumn,showpacs,preprintnumbers,amsmath,amssymb]{revtex4-1}

\usepackage{graphicx}
\usepackage{dcolumn}
\usepackage{bm}
\usepackage{natbib}
\usepackage[usenames,dvipsnames]{xcolor}
\usepackage{pifont}
\usepackage{relsize}
\definecolor{lightgreen}{rgb}{0,1,0}
\definecolor{darkgray}{gray}{0.20}

\begin{document}

\title{Enhanced reconstruction of weighted networks from strengths and degrees}

\author{Rossana Mastrandrea}
\affiliation{Institute of Economics and LEM, Scuola Superiore Sant'Anna, 56127 Pisa (Italy)\:}
\author{Tiziano Squartini}
\affiliation{Instituut-Lorentz for Theoretical Physics, University of Leiden,  2333 CA Leiden (The Netherlands)}
\author{Giorgio Fagiolo}
\affiliation{Institute of Economics and LEM, Scuola Superiore Sant'Anna, 56127 Pisa (Italy)}
\author{Diego Garlaschelli}
\affiliation{Instituut-Lorentz for Theoretical Physics, University of Leiden,  2333 CA Leiden (The Netherlands)\:}

\date{\today}

\begin{abstract}
Network topology plays a key role in many phenomena, from the spreading of diseases to that of financial crises. 
Whenever the whole structure of a network is unknown, one must resort to reconstruction methods that identify the least biased ensemble of networks consistent with the partial information available.
A challenging case, frequently encountered due to privacy issues in the analysis of interbank flows and Big Data, is when there is only local (node-specific) aggregate information available. For binary networks, the relevant ensemble is one where the degree (number of links) of each node is constrained to its observed value.
However, for weighted networks the problem is much more complicated.
While the na\"ive approach prescribes to constrain the strengths (total link weights) of all nodes, recent counter-intuitive results suggest that in weighted networks the degrees are often more informative than the strengths. 
This implies that the reconstruction of weighted networks would be significantly enhanced by the specification of both strengths and degrees, a computationally hard and bias-prone procedure. 
Here we solve this problem by introducing an analytical and unbiased maximum-entropy method that works in the shortest possible time and does not require the explicit generation of reconstructed samples.
We consider several real-world examples and show that, while the strengths alone give poor results, the additional knowledge of the degrees yields accurately reconstructed networks.
Information-theoretic criteria rigorously confirm that the degree sequence, as soon as it is non-trivial, is irreducible to the strength sequence. 
Our results have strong implications for the analysis of motifs and communities and whenever the reconstructed ensemble is required as a null model to detect higher-order patterns.
\end{abstract}



\maketitle

\section{Introduction}
A range of phenomena of critical importance, from the spread of infectious diseases to the diffusion of opinions and the propagation of financial crises, is highly sensitive to the topology of the underlying network that mediates the interactions \cite{barratsbook}.
This sensitivity implies that, whenever it is not possible to have a complete empirical knowledge of the network, one should make an optimal use of the partial information available and try to reconstruct the most likely network, or rather an ensemble of likely networks, in the least biased way.
In the Big Data era, this kind of problem is becoming more and more important given the ever-increasing availability of data that, for privacy issues, are often of aggregate nature \cite{bigdata1,bigdata2}.

Among the possible types of incomplete topological information (e.g. missing links, missing nodes, etc.), one of the most frequently encountered situations is when only a \emph{local} knowledge of the network is available \cite{mywtw,wells,bargigli,musmeci,guidonatphys,marsili}.
For instance, in binary networks knowing the \emph{number} of links (or `degree') of each node is typically much easier than knowing the \emph{identity} of all neighbours (the nodes at the other end of those links). 
Similarly, in weighted networks knowing the total intensity of links connected to each node (or `strength') is much easier than knowing the identity of all neighbours and  the intensity of all links separately.

A typical example is that of interbank networks, where it is relatively easy to know the total exposures of each bank, while privacy issues make it much more difficult to know \emph{who} is lending to whom, and \emph{how much} \cite{wells,bargigli,guidonatphys,marsili}.
Similarly, the Big Data phenomenon implies that a huge amount of information is continuously collected about individuals \cite{bigdata1,bigdata2}. In that case as well, privacy issues are becoming increasingly important, and methods that are able to give detailed predictions from aggregated data, while at the same time respecting the privacy of individuals, are therefore becoming more and more desirable.

Formally, network reconstruction can be regarded as a constrained entropy maximization problem, where the constraints represent the available information and the maximization of the entropy ensures that the reconstructed ensemble of networks is maximally random, given the enforced constraints \cite{newman_expo,mymethod}.
When the available information is just local, one only knows $O(N)$ quantities (e.g. the degrees of all nodes) instead of the total $O(N^2)$ ones (e.g. all entries of the adjacency matrix) fully describing the network.
This makes the network reconstruction problem very challenging, since the number of missing variables is still $O(N^2)$, i.e. of the same order of the total number.

Even when the real network is entirely known, it is often necessary to reconstruct the most likely network from local properties in order to have a benchmark (i.e. a \emph{null model}) to assess the statistical significance of any higher-order pattern, e.g. \emph{assortativity} 
\cite{vespy_w}, \emph{rich-club} effect \cite{myrichclub}, existence of \emph{network motifs} \cite{motifs,mytriadic} 
and \emph{communities} \cite{santo}.
Null models correctly filter out the intrinsic and unavoidable heterogeneity of nodes, e.g. the fact that more popular people naturally have a larger degree in social networks. 
The simplest and most extensively used null model is the \emph{Configuration Model} (CM), defined as an ensemble of random graphs with given \emph{degree sequence} (the vector of degrees of all nodes) \cite{newman_expo,mymethod}.
It was recently shown that, despite its conceptual simplicity, the CM already poses significant problems of \emph{bias}: it is very difficult to implement the model in such a way that each network in the reconstructed ensemble is assigned the correct probability and that the resulting ensemble-averaged expectations are unbiased \cite{mymethod,coolen}.
The problem of bias in the CM, or equivalently in the reconstruction of binary networks from local information, requires nontrivial solutions that have been proposed only recently \cite{mymethod,coolen,myjeic,mypre1}.
Once these solutions are appropriately implemented, many binary networks turn out to be reconstructed remarkably well from the knowledge of their degree sequence alone \cite{mymethod,myjeic,mypre1,mypre2}.
In other cases, the reconstructed network differs significantly from the real one, a result that is still very important as it reveals the presence of higher-order patterns that cannot be traced back to the degree sequence alone \cite{mymethod}.

In this paper we address the problem of the effective reconstruction, from local properties alone, of \emph{weighted} networks. 
We first show that, in contrast with what is generally believed, the reconstruction of weighted networks does not merely involve a one-to-one mapping of the corresponding methodology that works well for binary networks.
Specifically, inferring the structure of a weighted network only from the knowledge of its \emph{strength sequence} (the vector of strengths of all nodes) can lead to a very bad reconstruction, even for the networks that, at a binary level, can be reproduced extremely well from their degree sequence \cite{mymethod,myjeic,mypre2}.
We then conjecture that the reason is the fact that the knowledge of the strengths does not merely include or improve that of the degrees, since the binary information is completely lost once purely weighted quantities are measured.
This leads us to the expectation that the reconstruction of weighted networks would be significantly enhanced by the specification of both strengths and degrees. 
We therefore introduce an analytical and unbiased maximum-entropy technique to reconstruct unbiased ensembles of weighted networks from the knowledge of both strengths and degrees.
Our method directly provides, in the shortest possible time, the expected value of the desired reconstructed properties, in such a way that no explicit sampling of reconstructed graphs is required.
Moreover, being based on maximum-entropy distributions, our method is unbiased by construction.

In applying our enhanced method to several networks of different nature, we show that it leads to a significantly improved reconstruction, while remaining completely feasible since the required information is still local and the number of known variables is still $O(N)$.
We finally introduce rigorous information-theoretic criteria confirming that the joint specification of the strengths and degrees cannot be reduced to that of the strengths alone.
The resulting self-consistent picture is that the reconstruction of weighted networks is dramatically enhanced by the use of the irreducible set of joint degrees and strengths. 

Our results also have strong implications for the identification of higher-order patterns in real networks.
In particular, many of the observed properties that are unexplained by local weighted information do not necessarily call for non-local mechanisms as previously thought, since they turn out to be consistent with the enhanced, but still entirely local, information that includes both strengths and degrees.

\section{Na\"ive reconstruction of weighted networks}
Na\"ively, the most natural generalization of the CM to weighted networks is a reconstructed ensemble with given \emph{strength sequence}, and is sometimes referred to as the \emph{Weighted Configuration Model} (WCM) \cite{mymethod,serranoweighted1,serranoweighted2}.
The WCM is widely used both as a reconstruction method and as the most  important null model to detect communities.
In both cases, if $s_i$ denotes the strength of node $i$ and $N$ is the number of nodes, the expected weight of the link between nodes $i$ and $j$ predicted by the WCM is routinely written in the form
\begin{equation}
\langle w_{ij}\rangle = \frac{s_i s_j}{\sum_{m=1}^N s_m}
\label{eq:naive}
\end{equation}
or in a slightly different way if the network is directed (for simplicity, in this paper we will only consider undirected networks). 
For instance, the above expression represents one of the standard procedures to infer interbank linkages from the total exposures of individual banks \cite{wells}, or the fundamental null model used by most algorithms aimed at detecting densely connected \emph{communities} in weighted networks \cite{santo}.

Unfortunately, despite its widespread use, eq.(\ref{eq:naive}) is incorrect, and differs from the unbiased expression derived within a rigorous maximum-entropy approach \cite{mymethod,ginestra_wcm,mybosefermi}.
A simple signature of this inadequacy is the fact that, although eq.(\ref{eq:naive}) is treated as an expected value, there is no indication of the probability distribution from which it is derived. Therefore, it is impossible to derive the expected value of topological properties which are nonlinear functions of the weights (i.e. the weighted clustering coefficient that we will introduce later).
This problem has been solved only recently with the introduction of an analytical maximum-likelihood approach that leads to the correct expressions for the weight probability and any function of the expected weights \cite{mymethod}.

A more profound limitation of the WCM persists even when the model is correctly implemented. 
It should be noted that the motivation for using the WCM as the natural generalization of the CM to weighted networks is the implicit assumption that the strength is an improved node-specific property, superior to the degree because it encapsulates the extra information provided by link weights. 
However, recent counter-intuitive results have shown that, while the \emph{complete} knowledge of a weighted network conveys of course more information than the complete knowledge of just its binary projection, the strength sequence (which embodies only \emph{partial}, but weighted, information about the network) is often surprisingly less informative than the degree sequence (which embodies the corresponding partial, and even unweighted, piece of information) 
\cite{mymethod,myjeic,mypre1,mypre2}. In particular, several \emph{purely topological} properties of real weighted networks turn out to be reproduced much better by applying the CM to the binary projection of the graph, than by applying the WCM to the original weighted network \cite{mymethod,myjeic,mypre2}. The reason is that the strength sequence gives a very bad prediction of purely topological properties, and particularly the degrees: in fact, out of the many, possible ways to redistribute each node's strength among the $N-1$ other vertices irrespectively of the number of links being created, the WCM prefers those predicting much denser networks than the real ones \cite{mypre2}.

As a preliminary step of our analysis, we now confirm and extend these non-obvious findings to various networks of different nature.
We will later use the same networks to illustrate our enhanced method.
We consider the Italian Interbank network in year 1999 \cite{interbank}, three `classic' social networks collected in \cite{bk}, seven food webs from  \cite{foodweb}, and finally the aggregated World Trade Web (WTW) in year 2002 \cite{mypre2}.
The latter example, where nodes are world countries and links are their trade relationships (amount of imports and exports), is the system for which the role of strengths and degrees, when considered separately, has been studied in greatest detail \cite{myjeic,mypre1,mypre2}.
It therefore represents an ideal example to be included in our analysis.

\begin{figure*}[t]
\centerline{\includegraphics[width=.99\textwidth]{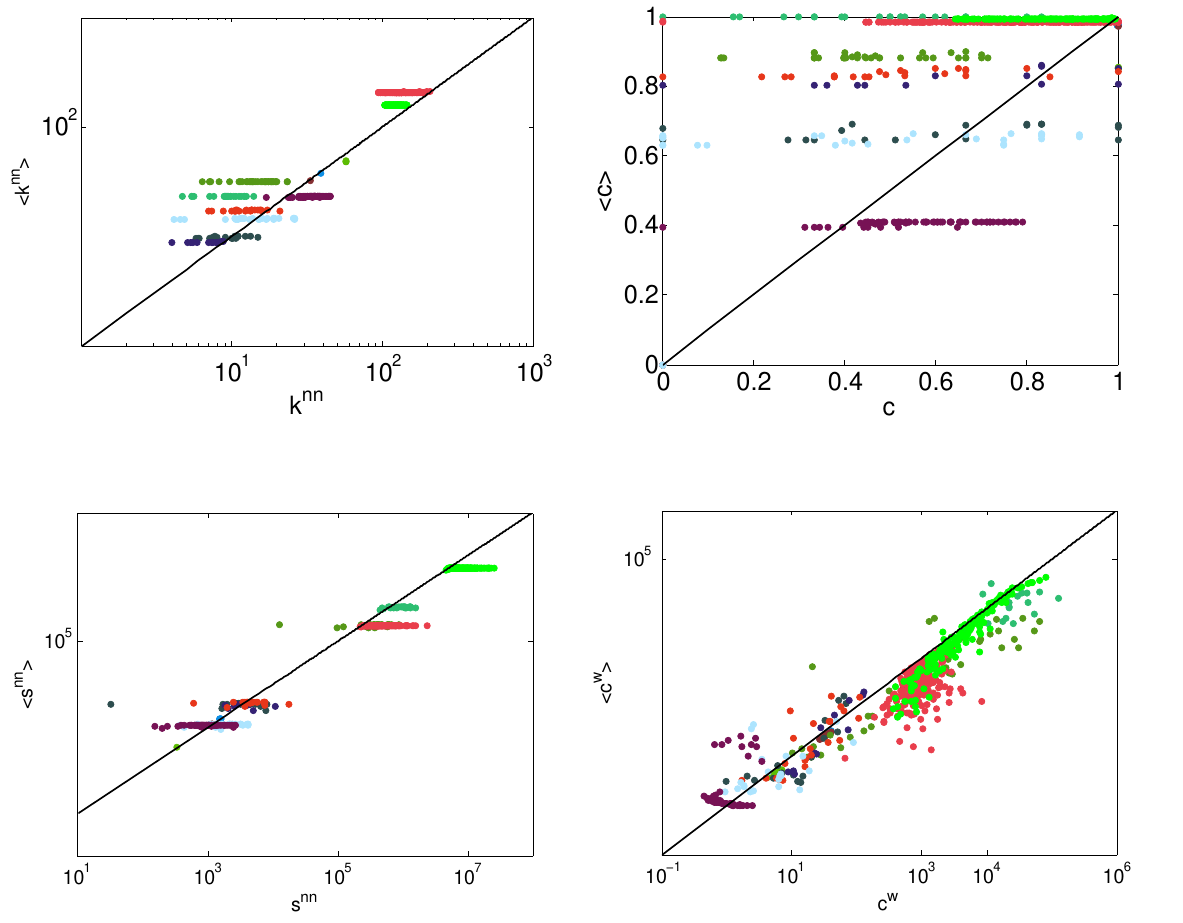}}
\caption{Na\"ive network reconstruction from node strengths (WCM), showing that purely weighted local properties are poorly informative.
In each panel we compare the reconstructed ($y$ axis) and real ($x$ axis) value of a node-specific network property, for all nodes of the following 12 networks: 
Office social network ($\textcolor{Cyan}{\mathlarger{\mathlarger{\mathlarger{\bullet}}}}$),  Research group social network ($\textcolor{Brown}{\mathlarger{\mathlarger{\mathlarger{\bullet}}}}$),  Fraternity social network ($\textcolor{LimeGreen}{\mathlarger{\mathlarger{\mathlarger{\bullet}}}}$), Maspalomas Lagoon food web ($\textcolor{SeaGreen}{\mathlarger{\mathlarger{\mathlarger{\bullet}}}}$),   Chesapeake Bay food web ($\textcolor{OliveGreen}{\mathlarger{\mathlarger{\mathlarger{\bullet}}}}$),  Crystal River (control) food web ($\textcolor{darkgray}{\mathlarger{\mathlarger{\mathlarger{\bullet}}}}$),  Crystal River food web ($\textcolor{BlueViolet}{\mathlarger{\mathlarger{\mathlarger{\bullet}}}}$),  Michigan Lake food web ($\textcolor{RedOrange}{\mathlarger{\mathlarger{\mathlarger{\bullet}}}}$),  Mondego Estuary food web ($\textcolor{blue!40!white}{\mathlarger{\mathlarger{\mathlarger{\bullet}}}}$),  Everglades Marshes food web ($\textcolor{Plum}{\mathlarger{\mathlarger{\mathlarger{\bullet}}}}$),  Italian Interbank network in year 1999 ($\textcolor{CarnationPink}{\mathlarger{\mathlarger{\mathlarger{\bullet}}}}$),  aggregated World Trade Web in year 2002 ($\textcolor{lightgreen}{\mathlarger{\mathlarger{\mathlarger{\bullet}}}}$).
Top left: average nearest neighbour degree ($k^{nn}_i$). 
Top right: binary clustering coefficient ($c_i$).
Bottom left: average nearest neighbour strength ($s^{nn}_i$). 
Bottom right: weighted clustering coefficient ($c^w_i$).  
\label{fig:wcm}}
\end{figure*}

From the above discussion, it is clear that in order to assess the performance of the network reconstruction method one should monitor not only the reconstructed  properties that depend entirely on link weights, but also those that depend on the binary topology. 
For this reason, in Fig.\ref{fig:wcm} we compare, for all networks in the sample, the empirical and reconstructed values of  various structural properties, including both purely topological properties and their weighted counterparts.
If the full weighted matrix is denoted by $\mathbf{W}$  (where $w_{ij}$ is the weight of the link between node $i$ and node $j$), the purely topological quantities are calculated on the binary projection $\mathbf{A}$ (adjacency matrix) of $\mathbf{W}$, with entries $a_{ij}=1$ if $w_{ij}>0$ and $a_{ij}=0$ if $w_{ij}=0$ (compactly, we can write $a_{ij}\equiv w^0_{ij}$ with the convention $0^0\equiv 0$).

The binary quantities we choose are the simplest non-local ones, i.e. those involving paths going two and three steps away from a node. 
The \emph{average nearest neighbor degree} (ANND), which is a measure of correlation between the degrees of adjacent nodes, is defined as
\begin{equation}
k^{nn}_i(\mathbf{W})\equiv\frac{\sum_{j\ne i}a_{ij}k_j}{k_i}=\frac{\sum_{j\ne i}\sum_{k\ne j}w^0_{ij}w^0_{jk}}{\sum_{j\ne i} w_{ij}^0}
\label{eq:annd}
\end{equation}
(where ${k_i=\sum_{j\ne i} a_{ij}=\sum_{j\ne i}w_{ij}^0}$)
and the \emph{clustering coefficient}, which measures the fraction of triangles around node $i$, is defined as
\begin{eqnarray}
c_i(\mathbf{W})
=\frac{\sum_{j\neq i}\sum_{k\ne i,j}w^0_{ij}w^0_{jk}w^0_{ki}}{\sum_{j\neq i}\sum_{k\ne i,j}w^0_{ij}w^0_{ki}}
\label{eq:cb}
\end{eqnarray}
The corresponding weighted quantities are the \emph{average nearest neighbor strength} (ANNS) \cite{mypre2} defined as
\begin{equation}{s}^{nn}_i(\mathbf{W})\equiv\frac{\sum_{j\ne i}a_{ij}{s}_j}{k_i}
=\frac{\sum_{j\ne i}\sum_{k\ne j}w^0_{ij}w_{jk}}{\sum_{j\ne i} w_{ij}^0}
\label{eq:anns}
\end{equation}
(where $s_i=\sum_{j\ne i} w_{ij}$)
and the \emph{weighted clustering coefficient} \cite{mypre2,giorgio_clustering} defined as
 \begin{eqnarray}
{c}_i^w(\mathbf{W})
=\frac{\sum_{j\neq i}\sum_{k\ne i,j}({w}_{ij}{w}_{jk}{w}_{ki})^{1/3}}{\sum_{j\neq i}\sum_{k\ne i,j}w^0_{ij}w^0_{ki}}
\label{eq:cw}
\end{eqnarray}

In the Four panels of fig. \ref{fig:wcm}, we show the measured value of the four quantities defined above, for all nodes and for all networks, and we compare it with the corresponding reconstructed value predicted by the WCM. The methodology used is described in refs. \cite{mymethod,mypre2} and briefly summarized later.
In this type of plot, every point is a node. Therefore the target of a good reconstruction method is that of placing all the points along the identity line.
By contrast, in most cases we find that the reconstructed values for all  nodes of a given network lie along  horizontal lines, i.e. they are nearly equal to each other and totally unrelated to the `target' real values. 

At this point, it should be noted that 
the typical interpretation of a result like the above one is that the reconstruction of networks from local node-specific information is intrinsically problematic, presumably because of higher-order mechanisms involved in the formation of real networks.
In fact, from the point of view of pattern detection, the WCM is often used as a null model to filter out the local heterogeneity of nodes in the detection of important higher-order properties such as communities \cite{mymethod,serranoweighted1,serranoweighted2}, 
thus interpreting the difference between real data and the WCM as an important signature of non-local patterns. 
Most community detection methods are indeed entirely based on this difference, and use it to define the so-called \emph{modularity} guiding the detection algorithm \cite{santo}.
However, as we show in the following, all the above results and the corresponding  interpretations are completely reversed if we consider an enhanced reconstruction method.

\section{The irreducibility conjecture}
In what follows, we propose a different interpretation of the above findings.
We conjecture (and rigorously prove later) that, in general, the poor reconstruction achieved by the WCM might be largely due to fact that the strength sequence discards purely topological information, and in particular the degrees. 
This hypothesis builds on previous results on the role of strengths and degrees in the WTW \cite{myjeic,mypre1,mypre2}. 
While, at a binary level, the assortativity and clustering properties of the WTW can be excellently reproduced by the CM \cite{mypre1}, the corresponding weighted quantities turn out to be very different from the ones predicted by the WCM on the basis of the strength sequence alone \cite{mypre2}.
These results are very robust and hold true over time, on different datasets, and for various resolutions of the WTW (i.e. for different levels of aggregation of traded commodities) \cite{myjeic,mypre1,mypre2}. 

\begin{figure*}[t]
\centerline{\includegraphics[width=.99\textwidth]{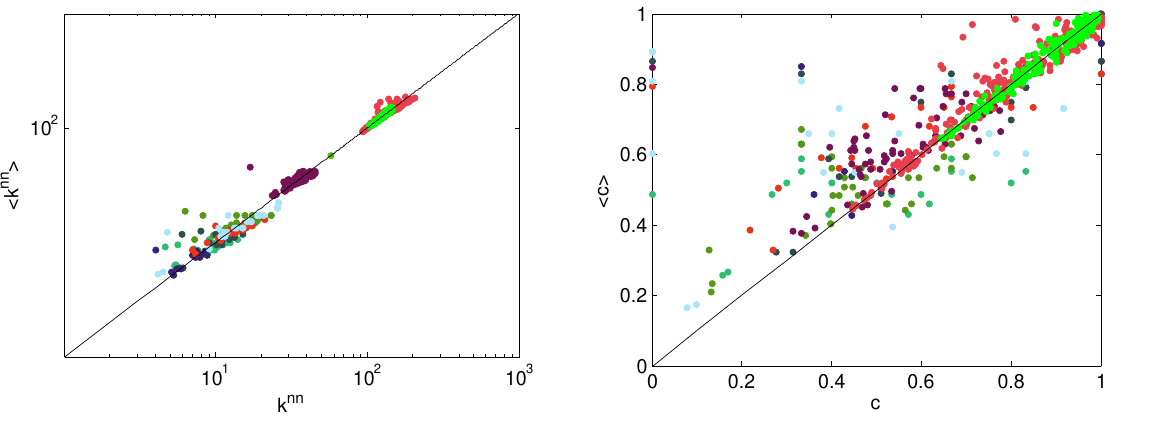}}
\caption{Reconstruction of the binary projection of the network from node degrees (CM), showing that purely binary local properties are significantly informative.
In each panel we compare the reconstructed ($y$ axis) and real ($x$ axis) value of a node-specific network property, for all nodes of the following 12 networks: 
Office social network ($\textcolor{Cyan}{\mathlarger{\mathlarger{\mathlarger{\bullet}}}}$),  Research group social network ($\textcolor{Brown}{\mathlarger{\mathlarger{\mathlarger{\bullet}}}}$),  Fraternity social network ($\textcolor{LimeGreen}{\mathlarger{\mathlarger{\mathlarger{\bullet}}}}$), Maspalomas Lagoon food web ($\textcolor{SeaGreen}{\mathlarger{\mathlarger{\mathlarger{\bullet}}}}$),   Chesapeake Bay food web ($\textcolor{OliveGreen}{\mathlarger{\mathlarger{\mathlarger{\bullet}}}}$),  Crystal River (control) food web ($\textcolor{darkgray}{\mathlarger{\mathlarger{\mathlarger{\bullet}}}}$),  Crystal River food web ($\textcolor{BlueViolet}{\mathlarger{\mathlarger{\mathlarger{\bullet}}}}$),  Michigan Lake food web ($\textcolor{RedOrange}{\mathlarger{\mathlarger{\mathlarger{\bullet}}}}$),  Mondego Estuary food web ($\textcolor{blue!40!white}{\mathlarger{\mathlarger{\mathlarger{\bullet}}}}$),  Everglades Marshes food web ($\textcolor{Plum}{\mathlarger{\mathlarger{\mathlarger{\bullet}}}}$),  Italian Interbank network in year 1999 ($\textcolor{CarnationPink}{\mathlarger{\mathlarger{\mathlarger{\bullet}}}}$),  aggregated World Trade Web in year 2002 ($\textcolor{lightgreen}{\mathlarger{\mathlarger{\mathlarger{\bullet}}}}$).
Left: average nearest neighbour degree ($k^{nn}_i$). 
Right: binary clustering coefficient ($c_i$).
\label{fig:bin}}
\end{figure*}

We know show that similar conclusions extend to all the networks in our analysis. 
While in Fig. \ref{fig:wcm} we have already illustrated the shortcomings of the WCM on several real networks, we have not inspected yet the performance of the CM when applied to the purely binary projection of the same networks.
In Fig.\ref{fig:bin} we compare the purely topological quantities considered above, i.e. the average nearest neighbor degree and the clustering coefficient of all nodes of our networks, with the prediction of the binary CM (thus obtained by only taking the degree sequence as input from the data \cite{mymethod}).
By comparing Fig. \ref{fig:bin} with the two upper panels of Fig. \ref{fig:wcm}, we clearly see that the CM is able to reconstruct the binary projection of the original networks much better that the WCM does, thus extending the results discussed in refs. \cite{myjeic,mypre1,mypre2} for the specific case of the WTW to a much broader class of real-world networks. 

Taken together, the results shown so far perfectly illustrate that the na\"ive expectation that quantities calculated on the original weighted network are \emph{per se} more informative than the corresponding quantities calculated on the binary projection is fundamentally incorrect. 
According to our conjecture, the degrees are instead to be considered a `fundamental' local structural property of weighted networks, irreducible to the knowledge of the strengths and thus at least as important as the latter.
Thus, the failure of the WCM might be due to the fact that, by discarding the degree sequence, the model is `violating' this irreducibility.

We should at this point clarify that by `irreducible' we do not refer to the \emph{numerical values} of strengths and degrees, but to the different \emph{functional roles} that the two quantities play in determining or constraining the network's structure. In fact, strengths and degrees are typically highly correlated in real networks \cite{vespy_w}, which means that we might be able to reasonably infer the values of one quantity from those of the other (in this sense, strengths and degrees are `reducible' to each other). 
However, what is of interest to us is a deeper form of irreducibility, encountered when the joint specification of strengths and degrees (even when the \emph{observed} numerical values of these quantities are perfectly correlated) 
\emph{constrains the network in a fundamentally different way} than the specification of only one of the two properties.
By the way, nothing guarantees that even a strong degree-strength correlation in the empirical network, i.e. a relation of the form $s_i=f(k_i)$, is preserved in an ensemble where only the strengths are controlled for, since for the ensemble averages one would generally get $\langle s_i\rangle\ne f(\langle k_i\rangle)$. 

\begin{figure*}[t]
\centerline{\includegraphics[width=.49\textwidth]{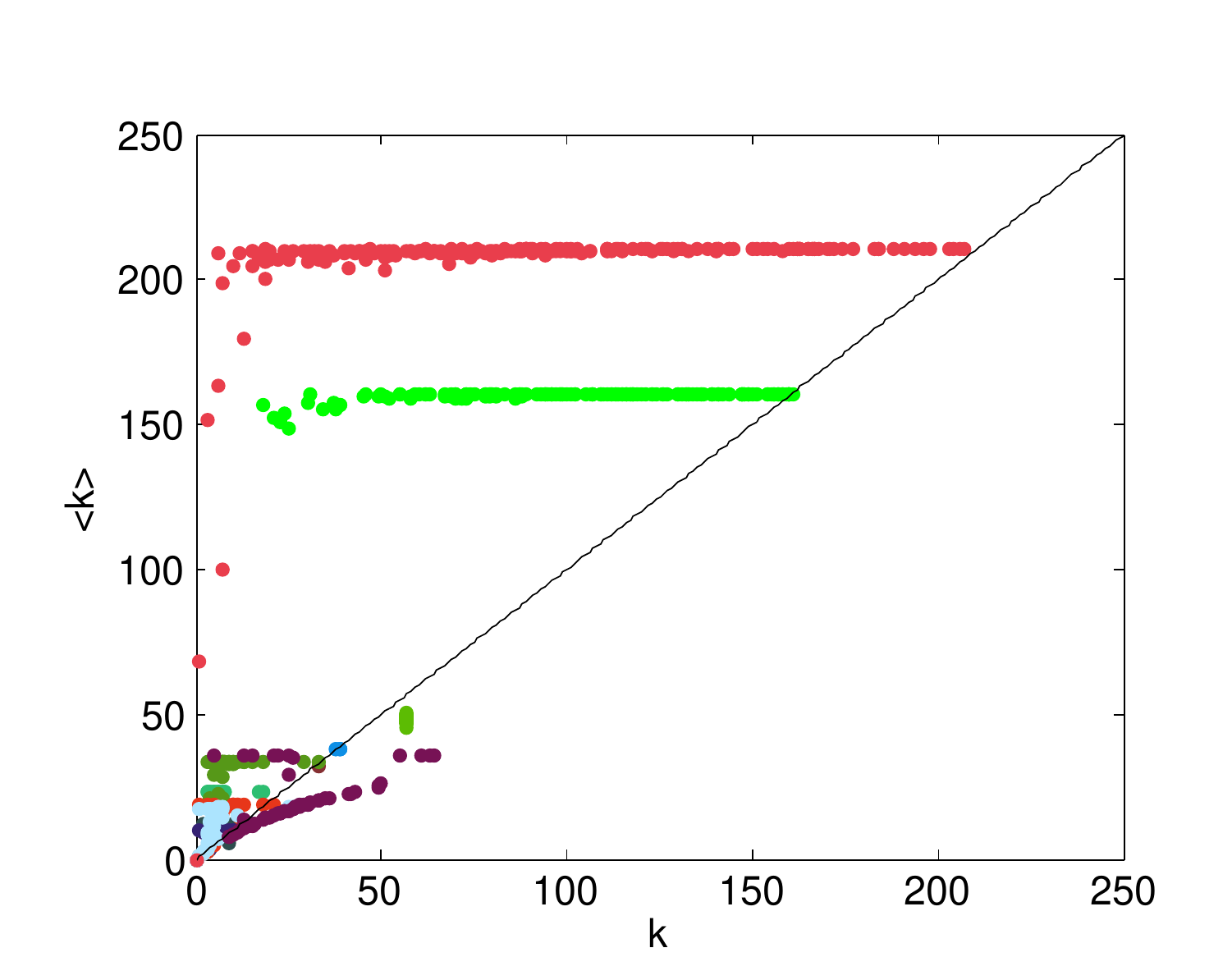}}
\caption{Reconstruction of node degrees from node strengths (WCM), showing that purely weighted local properties are poorly informative.
We compare the reconstructed ($y$ axis) and real ($x$ axis) value of the degree, for all nodes of the following 12 networks: 
Office social network ($\textcolor{Cyan}{\mathlarger{\mathlarger{\mathlarger{\bullet}}}}$),  Research group social network ($\textcolor{Brown}{\mathlarger{\mathlarger{\mathlarger{\bullet}}}}$),  Fraternity social network ($\textcolor{LimeGreen}{\mathlarger{\mathlarger{\mathlarger{\bullet}}}}$), Maspalomas Lagoon food web ($\textcolor{SeaGreen}{\mathlarger{\mathlarger{\mathlarger{\bullet}}}}$),   Chesapeake Bay food web ($\textcolor{OliveGreen}{\mathlarger{\mathlarger{\mathlarger{\bullet}}}}$),  Crystal River (control) food web ($\textcolor{darkgray}{\mathlarger{\mathlarger{\mathlarger{\bullet}}}}$),  Crystal River food web ($\textcolor{BlueViolet}{\mathlarger{\mathlarger{\mathlarger{\bullet}}}}$),  Michigan Lake food web ($\textcolor{RedOrange}{\mathlarger{\mathlarger{\mathlarger{\bullet}}}}$),  Mondego Estuary food web ($\textcolor{blue!40!white}{\mathlarger{\mathlarger{\mathlarger{\bullet}}}}$),  Everglades Marshes food web ($\textcolor{Plum}{\mathlarger{\mathlarger{\mathlarger{\bullet}}}}$),  Italian Interbank network in year 1999 ($\textcolor{CarnationPink}{\mathlarger{\mathlarger{\mathlarger{\bullet}}}}$),  aggregated World Trade Web in year 2002 ($\textcolor{lightgreen}{\mathlarger{\mathlarger{\mathlarger{\bullet}}}}$).
\label{fig:kk}}
\end{figure*}

The above line of reasoning leads us to expect that, in general, the WCM does not correctly reproduce the degree sequence of real networks. Again, this effect has been recently documented in the WTW \cite{myjeic,mypre2}. 
To provide further compelling evidence, in Fig.\ref{fig:kk} we compare the observed degrees of all nodes in our networks with the corresponding expectation under the WCM. 
We clearly see that most points are far from the identity line. Moreover, the majority of the reconstructed values lie along approximately constant lines, meaning that they are almost independent of the empirical values of the degree. These almost constant values are close to the maximum possible value $N-1$, indicating that the failure of the WCM is rooted in the fact that it incorrectly redistributes the observed strength of each node over too many edges, generally creating very dense (often almost completely connected) networks.
This result explains why, in Fig. \ref{fig:wcm}, the reconstructed values of $k^{nn}_i$, $c_i$ and $s^{nn}_i$ are approximately constant as well.
Indeed, it is easy to show that in an almost complete network these three quantities are necessarily nearly constant.

So, our conjecture leads us to the expectation that an enhanced reconstruction method (or null model) of weighted networks using purely local information should build on the simultaneous specification of strengths and degrees. 
Unfortunately, no satisfactory way to implement such method for the analysis of real networks has been proposed so far.
Moreover, no rigorous criterion has been defined to assess whether the introduction of the degree sequence as an additional constraint in the WCM is indeed non-redundant, i.e. not over-fitting the network.
It is therefore impossible, using the available techniques, to test the conjecture that the degrees are irreducible to the strengths.

In what follows, we fill both gaps by first defining a fast and unbiased approach to realize the enhanced network reconstruction method, and then introducing information-theoretic criteria to check \emph{a posteriori} whether the addition of degrees is non-redundant, confirming the irreducibility conjecture. 
Taken together, these two ingredients make the entire approach self-consistent and also show that the enhanced reconstructed ensemble should be considered as an improved null model of weighted networks with local properties. 

\section{Weighted networks with given strengths and degrees: the Enhanced Configuration Model}
For simplicity, we will refer to the ensemble of networks with given strengths and degrees as the `Enhanced Configuration Model' (ECM).
Early attempts to generate the ECM were either based on computational randomizations \cite{MCMwtw} or on theoretical arguments \cite{serranoweighted2}.
However, analytical calculations later showed that these approaches are statistically biased \cite{mybosefermi}. 
We now develop a maximum-entropy formalism that implements the ECM in an analytical, unbiased, and fast way. 
We only consider the case of undirected networks, although the generalization to the directed case is straightforward.
Formally, an ensemble of weighted networks with $N$ nodes can be characterized by a collection $\{\mathbf{W}\}$ of ${N\times N}$ matrices and by an appropriate probability ${P(\mathbf{W})}$ \cite{mybosefermi}.
On each network $\mathbf{W}$, 
the strength is defined as ${s_i(\mathbf{W})\equiv\sum_{j\ne i}w_{ij}}$ and the degree is defined as ${k_{i}(\mathbf{W})\equiv \sum_{j\ne i}w^0_{ij}}$.
We assume that each $w_{ij}$ is a non-negative integer number (again, with the convention $0^0=0$).

We start with a summary of useful analytical results that are already available \cite{mybosefermi}.
We look for a probability that, besides being normalized (${\sum_\mathbf{W}P(\mathbf{W})=1}$), 
ensures that the (expected) degree and strength of each node are both constrained, while leaving  the ensemble maximally random otherwise (thus not biasing the probability).
This is achieved by requiring that $P(\mathbf{W})$ maximizes 
Shannon's entropy $S\equiv-\sum_{\mathbf{W}}P(\mathbf{W})\ln P(\mathbf{W})$ with a constraint on the expected degree and strength sequences $\langle \vec{k}\rangle$, $\langle \vec{s}\rangle$.
The fundamental result \cite{mybosefermi} of this constrained maximization is the probability
\begin{equation}
P(\mathbf{W}|\vec{x},\vec{y})=\prod_{i<j}q_{ij}(w_{ij}|\vec{x},\vec{y})
\label{eq_P2}
\end{equation}
where $\vec{x}$ and $\vec{y}$ are two $N$-dimensional Lagrange multipliers controlling for the expected degrees and strengths respectively (with $x_i\ge 0$ and $0\le y_i<1$ $\forall i$), and 
\begin{equation}
q_{ij}(w|\vec{x},\vec{y})=\frac{(x_{i}x_{j})^{\Theta(w)}(y_{i}y_{j})^{w}(1-y_{i}y_{j})}{1-y_{i}y_{j}+x_{i}x_{j}y_{i}y_{j}}
\label{prob}
\end{equation}
is the probability that a link of weight $w$ exists between nodes $i$ and $j$. In the above expression, $\Theta(x)=1$ if $x>0$ and $\Theta(x)=0$ otherwise.
Note that $\sum_{w=0}^{+\infty} q_{ij}(w|\vec{x},\vec{y})=1$ $\forall i,j$.

Equation (\ref{prob}) defines the `mixed' Bose-Fermi distribution \cite{mybosefermi} where, due to the presence of $\Theta(w)$, the establishment of a link of unit weight between two nodes requires a different (higher if $x_i x_j>1$) `cost' than the reinforcement (by a unit of weight) of an already existing link. This feature is due to the presence of both binary and weighted constraints, and makes the ECM potentially very appropriate to model real networks.
However, as we mentioned, no method has been proposed so far to implement the ECM for empirical analyses.

To achieve this, we now apply the maximum-likelihood approach \cite{mymethod,mylikelihood} to the model. 
We consider a particular real weighted network $\mathbf{W}^*$, whose only degrees $k_i^*\equiv k_i(\mathbf{W}^*)$ and strengths $s_i^*\equiv s_i(\mathbf{W}^*)$ are known.
The log-likelihood of the ECM defined by eqs.(\ref{eq_P2}) and (\ref{prob}) reads
\begin{eqnarray}
&\mathcal{L}(\vec{x},\vec{y})\equiv
\ln P(\mathbf{W}^*|\vec{x},\vec{y})=\sum_{i<j}\ln q_{ij}(w^*_{ij}|\vec{x},\vec{y})=&\nonumber\\
&\sum_{i=1}^N \left(k_i^*\ln x_i +s_i^*\ln y_i\right)+\sum_{i<j}\ln \left(\frac{1-y_{i}y_{j}}
{1-y_{i}y_{j}+
x_{i}x_{j}y_{i}y_{j}}\right).&
\end{eqnarray}
We now look for the specific parameter values $\vec{x}^*,\vec{y}^*$ that maximize $\mathcal{L}(\vec{x},\vec{y})$.
A direct calculation, analogous to the simpler ones encountered in other null models \cite{mymethod,mylikelihood}, shows that $\vec{x}^*,\vec{y}^*$ can be obtained as the real solution to the following $2N$ coupled equations:
\begin{eqnarray}
\langle k_i\rangle&=&\sum_{j\ne i}\frac{x_i x_j y_i y_j}{1-y_i y_j+x_i x_j y_i y_j}=k_i^*\:\:\:\forall i\label{eq_k}\\
\langle s_i\rangle&=&\sum_{j\ne i}\frac{x_i x_j y_i y_j}{(1-y_i y_j)(1-y_i y_j+x_i x_j y_i y_j)}=s_i^*\:\:\:\forall i\quad\:\:\:
\label{eq_s}
\end{eqnarray}
Therefore, we find that the likelihood-maximizing values $\vec{x}^*,\vec{y}^*$ are precisely those ensuring that the expected degree and strength sequences coincide with the observed sequences $\vec{k}^*$ and  $\vec{s}^*$, thus solving our initial problem.

As we show below, the values $\vec{x}^*,\vec{y}^*$ contain all the information necessary to reconstruct the network. 
Thus the maximum-likelihood approach translates the time-consuming and bias-prone problem of the computational generation of several reconstructed networks  
into the much simpler problem of solving the $2N$ equations (\ref{eq_k}-\ref{eq_s}), or equivalently maximizing the function $\mathcal{L}(\vec{x},\vec{y})$ of $2N$ variables.
To find $\vec{x}^*$ and  $\vec{y}^*$, we chose to solve eqs.(\ref{eq_k}-\ref{eq_s}) using MatLab (the code is available on request).
Note that finding $\vec{x}^*$ and  $\vec{y}^*$ only requires the knowledge of the observed strengths and degrees, and not that of the entire network $\mathbf{W}^*$.
This is consistent of the fact that $\vec{k}^*$ and  $\vec{s}^*$ are the \emph{sufficient statistics} of the problem.
 
\section{Reconstructed properties}
Once the solutions $\vec{x}^*$ and  $\vec{y}^*$ are found, they can be used to obtain the reconstructed (ensemble-averaged) network properties analytically, with no need to actually measure such properties on any sampled network.  
Specifically, given a topological property $X(\mathbf{W})$ whose `true' (but in general unknown) value is $X^*\equiv X(\mathbf{W}^*)$, the reconstructed value can be calculated analytically as $\langle X\rangle\equiv\sum_\mathbf{W}X(\mathbf{W})P(\mathbf{W}|\vec{x}^*,\vec{y}^*)$. 
For most topological properties of interest, this involves calculating the expected product of (powers of) distinct matrix entries, which simply reads
\begin{equation}
\left\langle \sum_{i\ne j\ne k,\dots}w_{ij}^{\alpha}\cdot w_{jk}^{\beta}\cdot\:\dots\right\rangle=
\sum_{i\ne j\ne k,\dots}\langle w_{ij}^{\alpha}\rangle \cdot\langle w_{jk}^{\beta}\rangle\cdot\:\langle\dots\rangle
\label{eq:product}
\end{equation}
with the generic term given by
\begin{equation}
\langle w_{ij}^{\gamma}\rangle=\!
\sum_{w=0}^{+\infty}\! w^{\gamma}q_{ij}(w|\vec{x}^*,\vec{y}^*)=\!
\frac{x^*_{i}x^*_{j}(1-y^*_{i}y^*_{j})\mbox{Li}_{-\gamma}(y^*_{i}y^*_{j})}{1-y^*_{i}y^*_{j}+x^*_{i}x^*_{j}y^*_{i}y^*_{j}}
\label{eq:li}
\end{equation}
where $\mbox{Li}_{n}(z)\equiv\sum_{l=1}^{+\infty}z^l/l^n$ is the $n$th polylogarithm of $z$.
The simplest and most useful cases $\gamma =1$ and $\gamma=0$ yield the expected weight $\langle w_{ij}\rangle$ and the connection probability ${p_{ij}=\langle \Theta(w_{ij})\rangle=\langle w_{ij}^0\rangle}$, respectively.
Therefore the reconstructed value $\langle X\rangle$ can be calculated in the same time as that required to calculate the real (if known) value $X(\mathbf{W}^*)$ (i.e. the shortest possible time), by simply replacing $w_{ij}^{\gamma}$ with $\langle w_{ij}^{\gamma}\rangle$ in the definition of $X(\mathbf{W})$.

\section{Enhanced reconstruction of real weighted networks}
\begin{figure*}[t]
\centerline{\includegraphics[width=.99\textwidth]{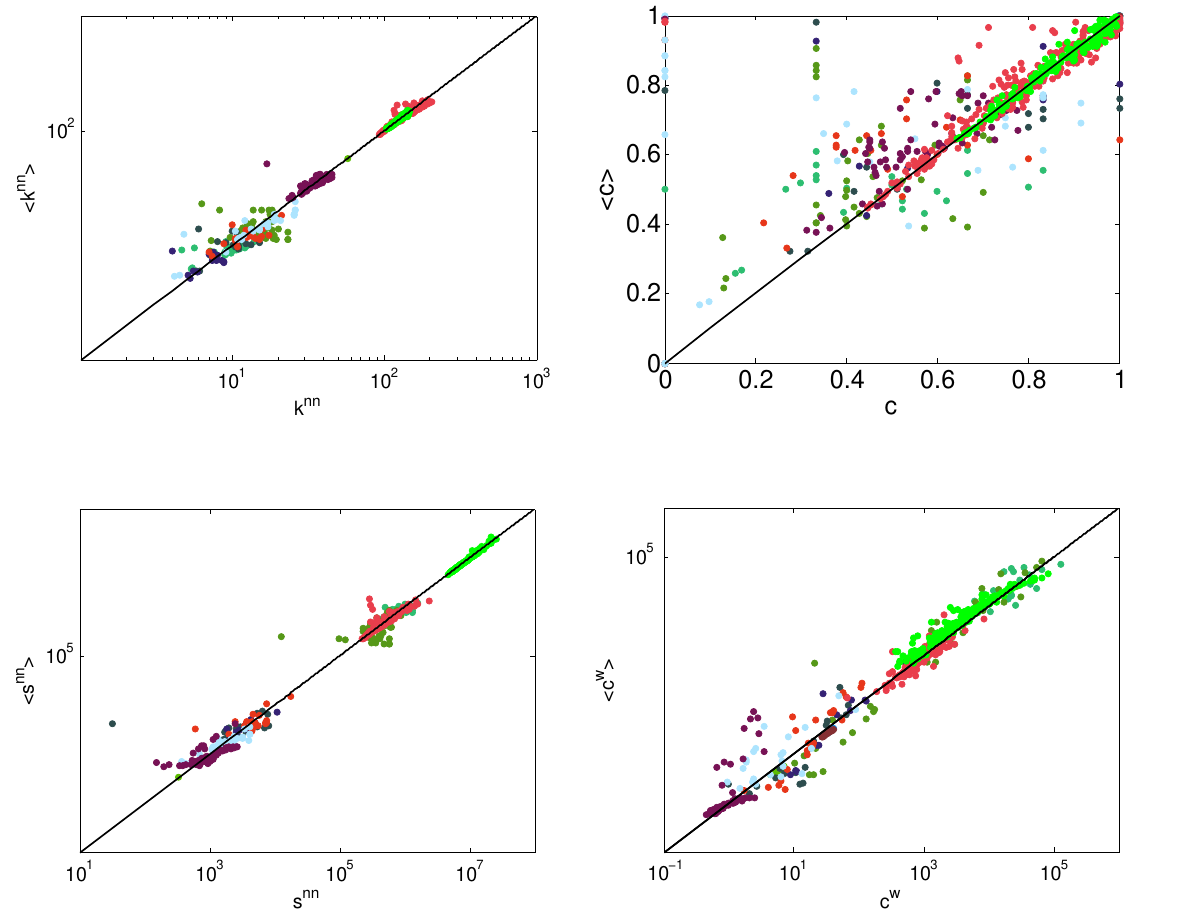}}
\caption{Enhanced network reconstruction from strengths and degrees (ECM), showing dramatic improvements over the standard approach shown previously in Fig. \ref{fig:wcm}.
In each panel we compare the reconstructed ($y$ axis) and real ($x$ axis) value of a node-specific network property, for all nodes of the following 12 networks: 
Office social network ($\textcolor{Cyan}{\mathlarger{\mathlarger{\mathlarger{\bullet}}}}$),  Research group social network ($\textcolor{Brown}{\mathlarger{\mathlarger{\mathlarger{\bullet}}}}$),  Fraternity social network ($\textcolor{LimeGreen}{\mathlarger{\mathlarger{\mathlarger{\bullet}}}}$), Maspalomas Lagoon food web ($\textcolor{SeaGreen}{\mathlarger{\mathlarger{\mathlarger{\bullet}}}}$),   Chesapeake Bay food web ($\textcolor{OliveGreen}{\mathlarger{\mathlarger{\mathlarger{\bullet}}}}$),  Crystal River (control) food web ($\textcolor{darkgray}{\mathlarger{\mathlarger{\mathlarger{\bullet}}}}$),  Crystal River food web ($\textcolor{BlueViolet}{\mathlarger{\mathlarger{\mathlarger{\bullet}}}}$),  Michigan Lake food web ($\textcolor{RedOrange}{\mathlarger{\mathlarger{\mathlarger{\bullet}}}}$),  Mondego Estuary food web ($\textcolor{blue!40!white}{\mathlarger{\mathlarger{\mathlarger{\bullet}}}}$),  Everglades Marshes food web ($\textcolor{Plum}{\mathlarger{\mathlarger{\mathlarger{\bullet}}}}$),  Italian Interbank network in year 1999 ($\textcolor{CarnationPink}{\mathlarger{\mathlarger{\mathlarger{\bullet}}}}$),  aggregated World Trade Web in year 2002 ($\textcolor{lightgreen}{\mathlarger{\mathlarger{\mathlarger{\bullet}}}}$).
Top left: average nearest neighbour degree ($k^{nn}_i$). 
Top right: binary clustering coefficient ($c_i$).
Bottom left: average nearest neighbour strength ($s^{nn}_i$). 
Bottom right: weighted clustering coefficient ($c^w_i$).  
\label{fig:mcm}}
\end{figure*}

We can now apply our general methodology to the reconstruction of real-world networks.
We consider again the assortativity and clustering properties defined in eqs.(\ref{eq:annd})-(\ref{eq:cw}).
The reconstructed value of all the above quantities can be simply obtained by replacing $w_{ij}^{\gamma}$ with $\langle w_{ij}^{\gamma}\rangle$ in such equations.
The result is illustrated in fig. \ref{fig:mcm} for all the networks shown previously in fig. \ref{fig:wcm}.
We clearly see that our enhanced method achieves a dramatic improvement over the standard approach.
Now most points lie in the vicinity of the identity, meaning that our method is able to successfully reconstruct, for each vertex, the structure of the network two and three steps away from it.
Note that the noisiest property is the binary clustering coefficient; however if we compare our results with the na\"ive ones we find that the improvement achieved for this quantity is perhaps the most significant one.

The above findings completely reverse the conclusions one would draw from the interpretation of the na\"ive results. 
First, network reconstruction from purely local properties is now shown to be possible to a highly satisfactory level, at least for the networks considered here.
Second, the assortativity and clustering properties of these networks turn out to be well explained by purely local, even if augmented, properties.  
So, there is no need to invoke non-local mechanisms in order to explain such properties in these networks.
We similarly expect that, if one considers the ECM as an improved null model to detect communities or other higher-order patterns, the result will be dramatically different from what is routinely obtained by using the WCM prediction in the definition of the modularity \cite{santo}. 
All these considerations suggests that, besides representing an improved reconstruction method, the ECM has the potential to become a nontrivial tool as a null model of networks with local constraints.

\section{Information-theoretic tests of irreducibility}
So far, we have assessed the superiority of our enhanced reconstruction method on the basis of its increased accuracy, with respect to the na\"ive approach, in reproducing the four `target' properties shown in fig. \ref{fig:mcm}.
We now confirm these results using a rigorous goodness-of-fit approach that compares the performance of the WCM and ECM in reproducing the \emph{whole} network.
At the same time, this approach will automatically allow us to test our initial conjecture that the degrees are irreducible to the strengths.
Indeed, both problems can be equivalently stated within a model selection framework, where one is interested in determining not only which of the two models achieves the best fit to the data, but also whether the introduction of the degrees as extra parameters in the ECM is really non-redundant, i.e. whether it does not over-fit the network.

To start with, we need to compare the likelihood of the ordinary WCM with that of ECM. 
Note that the WCM can be obtained as a particular case of the ECM by setting $\vec{x}=\vec{1}$ (where $x_i=1$ $\forall i$), i.e. by `swicthing off' the parameters controlling for the degrees. 
The log-likelihood of the WCM is therefore the reduced function $\mathcal{L}(\vec{1},\vec{y})$ of $N$ variables, and is maximized by a new vector $\vec{y}^{**}\ne \vec{y}^{*}$ which is also the solution of eq.(\ref{eq_s}) with $\vec{x}=\vec{1}$. In the WCM, eq.(\ref{eq_k}) no longer plays a role.
The predictions of the WCM are still obtained as in eqs.(\ref{eq:product}) and (\ref{eq:li}), by replacing $x^*_i$ with $1$ and $y^*_i$ with $y_i^{**}$ in the latter.
This is how the reconstructed properties plotted in Fig.\ref{fig:wcm} were computed.

Now, if we simply compare the maximized likelihoods of the two reconstruction methods, we trivially obtain $\mathcal{L}(\vec{x}^*,\vec{y}^*)\ge\mathcal{L}(\vec{1},\vec{y}^{**})$ since the ECM always improves the fit to the real network $\mathbf{W}^*$, given that it 
includes the WCM as a particular case and has extra parameters.
However, statistical and information-theoretic criteria exist \cite{akaike} to assess whether the increased accuracy of a model with more parameters is a result of over-fitting, in which case a more parsimonious model should be preferred.
The most popular choices are the Likelihood-ratio test (LRT), Akaike's Information Criterion (AIC), corrected Akaike's Information Criterion (AICc) and the Bayesian Information Criterion (BIC) \cite{akaike}.
These tests rigorously implement the idea that the optimal trade-off between accuracy and parsimony is achieved by discounting the number of free parameters from the maximized likelihood, and they differ in the way this discount is quantitatively implemented.
The simplest criterion is AIC, which (for our two competing null models) is defined as \begin{eqnarray}
\mbox{AIC}_{ECM}&\equiv& -2\mathcal{L}(\vec{x}^*,\vec{y}^*)+4N\\
\mbox{AIC}_{WCM}&\equiv& -2\mathcal{L}(\vec{1},\vec{y}^{**})+2N
\end{eqnarray}
The optimal model to be choose is the one minimizing AIC; however, if the difference between the AIC values is small, the two models will still be comparable.
A correct quantitative criterion is given by the so-called AIC weights \cite{akaike}, which in our case read
\begin{eqnarray}
w^{AIC}_{ECM}&\equiv&\frac{e^{-AIC_{ECM}/2}}{e^{-AIC_{ECM}/2}+e^{-AIC_{WCM}/2}}
\label{eq_wmcm}\\
w^{AIC}_{WCM}&\equiv& 1-w^{AIC}_{ECM}\label{eq_wwcm}
\end{eqnarray}
and quantify the weight of evidence in favour of each model, i.e. the probability that the model is the best one.

\begin{table}[t]
\caption{AIC weights for the considered null models (AICc and BIC weights give exactly the same results).\label{tab}}
\begin{tabular}{lcc}
\hline
$\mbox{\bf Network}$ & $ \mathbf{w_{WCM}^{AIC}} $ & $ \mathbf{w_{ECM}^{AIC}} $\\
\hline
\hline
$\textcolor{Cyan}{\mathlarger{\mathlarger{\mathlarger{\bullet}}}}$ Office social network \cite{bk} & $1$ & $0$\\
\hline
$\textcolor{Brown}{\mathlarger{\mathlarger{\mathlarger{\bullet}}}}$ Research group social network\cite{bk} & $1$ & $0$\\
\hline
$\textcolor{LimeGreen}{\mathlarger{\mathlarger{\mathlarger{\bullet}}}}$ Fraternity social network \cite{bk} & $0$ & $1$\\
\hline
$\textcolor{SeaGreen}{\mathlarger{\mathlarger{\mathlarger{\bullet}}}}$  Maspalomas Lagoon food web \cite{foodweb} & $0$ & $1$\\
\hline
$\textcolor{OliveGreen}{\mathlarger{\mathlarger{\mathlarger{\bullet}}}}$ Chesapeake Bay food web \cite{foodweb} & $0$ & $1$\\
\hline
$\textcolor{darkgray}{\mathlarger{\mathlarger{\mathlarger{\bullet}}}}$ Crystal River (control) food web \cite{foodweb} & $0$ & $1$\\
\hline
$\textcolor{BlueViolet}{\mathlarger{\mathlarger{\mathlarger{\bullet}}}}$ Crystal River food web \cite{foodweb} & $0$ & $1$\\
\hline
$\textcolor{RedOrange}{\mathlarger{\mathlarger{\mathlarger{\bullet}}}}$  Michigan Lake food web \cite{foodweb}  & $0$ & $1$\\
\hline
$\textcolor{blue!40!white}{\mathlarger{\mathlarger{\mathlarger{\bullet}}}}$ Mondego Estuary food web \cite{foodweb} & $0$ & $1$\\
\hline
$\textcolor{Plum}{\mathlarger{\mathlarger{\mathlarger{\bullet}}}}$ Everglades Marshes food web \cite{foodweb}  & $0$ & $1$\\
\hline
$\textcolor{CarnationPink}{\mathlarger{\mathlarger{\mathlarger{\bullet}}}}$ Italian interbank network (1999) \cite{interbank}  & $0$ & $1$\\
\hline
$\textcolor{lightgreen}{\mathlarger{\mathlarger{\mathlarger{\bullet}}}}$ World Trade Web (2000)\cite{mypre2} & $0$ & $1$\\
\hline
\end{tabular}
\end{table}

The AIC weights of the two reconstruction methods are shown in table \ref{tab} for all networks. 
We see that, apart from two social networks, the enhanced method is always superior to the na\"ive one, and achieves unit probabilty (within machine precision) of being the best among the two models.
A closer inspection of the two networks for which the opposite result holds reveals that they are (almost) fully connected.
This explains why the specification of the degree sequence, which in this case is close to the almost fully connected prediction of the WCM, is redundant for these networks.
In such cases, the relevant local constraints effectively reduce to the strength sequence, so the `standard' WCM is preferable.
Our method correctly indentifies this situation.
However, as soon as the topology is nontrivial (as in most real-world networks), the local constraints are irreducible to the strength sequence alone and the degrees must be separately specified in order to achieve a better reconstruction.
We should therefore expect that, for the vast majority of real-world networks, the degree sequence is irreducible to the strength sequence.
In such cases, the inclusion of degrees in our enhanced method is non-redundant, explaining why our method retrieves significantly more information. 

We also used AICc, that corrects for small samples, and BIC, that puts a higher penalty on the number of parameters \cite{akaike}.
Starting from the values of AICc and BIC, the corresponding weights are computed in analogy with eqs.(\ref{eq_wmcm}) and (\ref{eq_wwcm}).
We found that both the AICc and BIC weights are identical to the AIC ones (within machine precision) for all networks in our samples. 
Moreover, the LRT response is the same of AIC, AICc and BIC, at both 5\% and 1\% significance levels.

\section{Conclusions}
Motivated by recent findings suggesting that the properties calculated on the binary projection of real networks can be surprisingly more informative than the same properties calculated on the original weighted networks, in this work we have introduced an improved, fast and unbiased method to reconstruct weighted networks from the joint set of strengths and degrees.
We compared our enhanced method (ECM) with the simpler one that na\"ively uses only the strength sequence to reconstruct the network (WCM).

We confirmed an extremely bad agreement between real network properties and their WCM-reconstructed counterparts, implying that the strength sequence is in general uninformative about the higher-order properties of the network.
The typical interpretation of this result is that the network is shaped by non-local mechanisms, irreducible to local formation rules.
By contrast, we showed that the ECM provides accurate reconstructed properties, clearly outperforming the na\"ive approach and 
indicating that the combination of strengths and degrees is extremely informative.
In other words, the real networks in our analysis turned out to be typical members of the ECM ensemble and not of the WCM ensemble.
This has important consequences for important problems like the reconstruction of interbank linkages from bank-specific information: the analysis of the interbank network considered here shows that our approach is accurate while 
the standard one is uninformative.

Moreover, information-theoretic criteria confirmed that the inclusion of the degrees as additional constraints is non-redundant and does not `overfit' the network.
So strengths and degrees turn out to jointly represent an irreducible piece of local information for most real networks. 
An important consequence is that our ECM should be regarded as a more appropriate, and still parsimonious, null model of weighted networks with local constraints.
The agreement of this stricter null model with the networks in our sample implies that the higher-order properties considered here are well explained by local constraints, thus completely inverting the conclusions following from the use of the na\"ive approach.

\begin{acknowledgments}
D.G. acknowledges support from the Dutch Econophysics Foundation (Stichting Econophysics, Leiden, the Netherlands) with funds from beneficiaries of Duyfken Trading Knowledge BV, Amsterdam, the Netherlands. 
This work was also supported by the EU project MULTIPLEX (contract 317532) and the Netherlands Organization for Scientific Research (NWO/OCW).

G.F. gratefully acknowledges financial support received by the research project ``The international trade network: empirical analyses and theoretical models'' funded by the Italian Ministry of Education, University and Research (Scientific Research Programs of National Relevance 2009).
\end{acknowledgments}

\bibliographystyle{apsrev4-1}

\end{document}